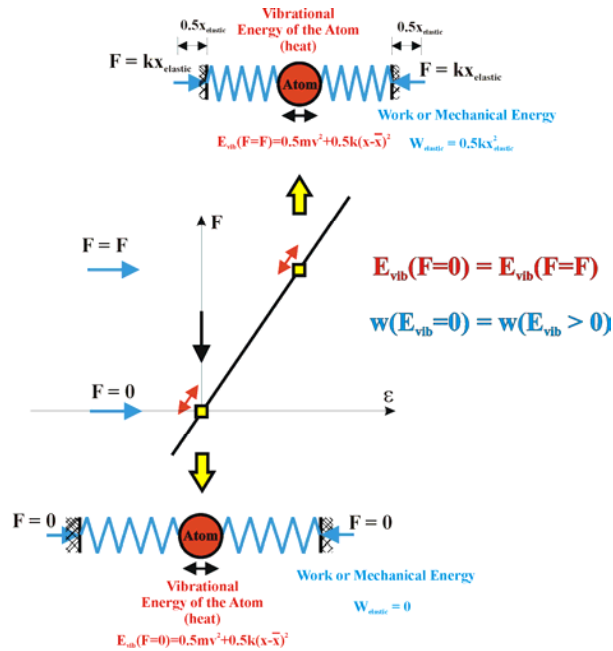

One dimensional vibration model demonstrating the independence of vibration (thermal) and elastic deformation (work). The force F refers to the maximum kinetic energy. In real systems limited energy exchange between heat and work (thermo-elastic coupling) is exist but the majority of heat and work is not transferable into each other in solid phase in the elastic domain. Thus Joule's postulation regarding to the mechanical equivalency of heat is not universal and applicable only if the energies are stored by the same physical process.

# The First Law of Thermodynamics

# and the Thermodynamic Description of Elastic Solids

Jozsef Garai

*Department of Mechanical and Materials Engineering, Florida International University, Miami, USA*

Historically, the thermodynamic behavior of gasses was described first and the derived equations were adapted to solids. It is suggested that the current thermodynamic description of solid phase is still incomplete because the isothermal work done on or by the system is not counted in the internal energy. It is also suggested that the isobaric work should not be deducted from the internal energy because the system does not do work when it expands. Further more it is suggested that Joule's postulate regarding the mechanical equivalency of heat "the first law of thermodynamics" is not universal and not applicable to elastic solids. The equations for the proposed thermodynamic description of solids are derived and tested by calculating the internal energies of the system using the equation of state of MgO. The agreement with theory is good.

## 1. Introduction

The Joule paddle-wheel experiment demonstrated that the mechanical work required for rotating the wheel is transferred to heat [1]. Based on this experiment, Joule postulated the mechanical equivalency of heat in the year 1845. This statement is known as the first law of thermodynamics and usually expressed as the sum of the differentials of the heat [q] and work [w]

$$dU = \delta q + \delta w \qquad (1)$$

where U is the internal energy of the system. The expression of Eq. (1) is also interpreted as a statement of energy conservation, which is one of the fundamental laws of nature. The energy conservation in Eq. (1) is accepted universally; however, the mechanical equivalency of heat is questioned. Joule's experiment demonstrated only the one way transformation of mechanical work to heat. His postulation for the universality of the mechanical equivalency of heat has never been proven experimentally. The correctness of Eq. (1) is questioned ever since its postulation [ex. 2, 3]. Articles suggesting possible flaws in the conceptual bases of the classical thermodynamics are continuously published [4, 5 and ref. therein]; however, these objections are unjustly ignored in the contemporary literature.

Equation (1) can be expressed in a various ways. Assuming a quasi-static process and introducing entropy [S] [6] as:

$$dS \equiv \frac{dq}{T} \qquad (2)$$

and the expansion work against hydrostatic pressure [p]

$$dw = -pd(V) \qquad (3)$$

where V is the volume and the principal stresses are $p = \sigma_1 = \sigma_2 = \sigma_3$. Substituting Eqs. (2) and (3) into Eq. (1) gives the differentials of the internal energy as:

$$dU = TdS - pdV. \qquad (4)$$

The two potentials are the pressure

$$p = -\left(\frac{\partial U}{\partial V}\right)_S, \qquad (5)$$

and the temperature [T]

$$T = \left(\frac{\partial U}{\partial S}\right)_V. \qquad (6)$$

Sometimes it is more convenient to measure the volume and the temperature and express the internal energy as:

$$dU = \left(\frac{\partial U}{\partial T}\right)_V dT + \left(\frac{\partial U}{\partial V}\right)_T dV \qquad (7)$$

The temperature derivative of the internal energy is the heat capacity at constant volume $[C_V]$

$$C_V = \left(\frac{\partial U}{\partial T}\right)_V, \qquad (8)$$

which is usually given in molar units $[c_V]$

$$c_V = \frac{1}{n}\left(\frac{\partial U}{\partial T}\right)_V, \qquad (9)$$

where n is the number of moles. The volume derivative of the internal energy can be written as:

$$\left(\frac{\partial U}{\partial V}\right)_T = T\left(\frac{\partial p}{\partial T}\right)_V - p \qquad (10)$$

where

$$\left(\frac{\partial p}{\partial T}\right)_V = \alpha B_T. \qquad (11)$$

The two parameters, the volume coefficient of expansion $[\alpha]$ and the isothermal bulk modulus $[B_T]$ are defined as:

$$\alpha \equiv \frac{1}{V}\left(\frac{\partial V}{\partial T}\right)_P \qquad (12)$$

and

$$B_T \equiv -V\left(\frac{\partial p}{\partial V}\right)_T. \qquad (13)$$

It is assumed that the solid is homogeneous, isotropic, and non-viscous. The internal energy is then:

$$dU = nc_V dT - [p - T\alpha B_T]dV. \qquad (14)$$

Introducing the thermal pressure $[p^{th}]$

$$p^{th} = \int_{T=0}^{T}\left(\frac{\partial p}{\partial T}\right)_V dT = \int_{T=0}^{T}\alpha_V K_T dT \qquad (15)$$

gives the expression of the first law of thermodynamics in the solid phase as:

$$dU = nc_V dT - [p - p^{th}]dV. \qquad (16)$$

In this study the long outstanding debate regarding the mechanical equivalency of heat and the fundamental equations of thermodynamics Eq. (1), (4) and (16) are revisited and investigated in solid phase in the elastic domain.

## 2. Internal energy

Historically, the thermodynamic behavior of gasses was described first and the derived equations were adapted to solids. This adoption seems to be incomplete because the differential of the mechanical work [Eq. (3)] and the differential of the internal energy [Eq. (16)] contain only the work performed at isobaric conditions. Work can also be done at isothermal condition which should be included in the internal energy of the system.

$$dw_T = (p + dp)d[V]_T \qquad (17)$$

The pressure reduces the value of the heat capacity [7]. If a system is heated up at different pressures to the same temperature then at higher pressures always requires less heat despite the system does more isobaric work on the surroundings.

$$q_{p=0,T} > q_{p,T} \quad \text{contrarily} \quad w_{p=0,T} < w_{p,T} \qquad (18)$$

This reverse relationship between heat and isobaric work indicates that the isobaric work is independent from heat. This independence can be explained in the following way. Let assume that the system is surrounded by the same substance as the system itself. In this case when the system is heated up no isobaric work is done by the system because the surrounding expands in the same manner. Thus the isobaric pressure work should not be deducted from the energy budget of heat. The same conclusion can be reached from Eq. (16) which gives non zero isobaric work at zero pressure when no work on or by the surrounding is done. The differential of the internal energy should be given then as:

$$dU^s = \delta q_p + \delta w_T. \qquad (19)$$

where superscript s is used for solid phase. The differential of the internal energy is then

$$dU^s = \left(\frac{\partial q}{\partial T}\right)_p dT + \left(\frac{\partial w_T}{\partial p}\right)_T dp. \qquad (20)$$

The work term in Eq. (19) and (20) represents only the isothermal work because the isobaric work is included in the heat term. The differential of the isothermal work at constant pressure is zero

$$d(w_T)_p = 0. \qquad (21)$$

therefore

$$\left(\frac{\partial q_p}{\partial T}\right) = \left(\frac{\partial U^s}{\partial T}\right)_p. \qquad (22)$$

The differential of the heat at constant temperature is zero

$$d(q_p)_T = 0. \qquad (23)$$

therefore
$$\left(\frac{\partial w_T}{\partial p}\right) = \left(\frac{\partial U^s}{\partial p}\right)_T. \quad (24)$$

The differential of the internal energy then can be written as:
$$dU^s = \left(\frac{\partial U^s}{\partial T}\right)_p dT + \left(\frac{\partial U^s}{\partial p}\right)_T dp. \quad (25)$$

The differential of the isothermal volume $d[V]_T$ is:
$$d[V]_T = \left(\frac{\partial V}{\partial p}\right)_T dp. \quad (26)$$

and
$$dp = \left(\frac{\partial p}{\partial V}\right)_T d[V]_T. \quad (27)$$

Substituting this expression into Eq. (25) gives the differential of the internal energy as:
$$dU^s = \left(\frac{\partial U^s}{\partial T}\right)_p dT + \left(\frac{\partial U^s}{\partial p}\right)_T \left(\frac{\partial p}{\partial V}\right)_T d[V]_T = \left(\frac{\partial U^s}{\partial T}\right)_p dT + \left(\frac{\partial U^s}{\partial V}\right)_T d[V]_T. \quad (28)$$

Eventhough the sum of the heat and work is path independent the ratio between these energies changes along the path because thermoelastic coupling occurs. The pressure and temperature effect on the heat and the work parts of the internal energy can be taken into consideration as:
$$\left(\frac{\partial U^s}{\partial T}\right)_p dT = \left(\frac{\partial U^s}{\partial T}\right)_{p=0} dT + \left(\frac{\partial U^s_{p=0}}{\partial p}\right)_T dp, \quad (29)$$

and
$$\left(\frac{\partial U^s}{\partial V}\right)_T d[V]_T = \left(\frac{\partial U^s}{\partial V}\right)_{T=0} dV + \left(\frac{\partial U^s_{T=0}}{\partial T}\right)_p dT. \quad (30)$$

The differential of the internal energy then can be given as:
$$dU^s = \left(\frac{\partial U^s}{\partial T}\right)_{p=0} dT + \left(\frac{\partial U^s_{p=0}}{\partial p}\right)_T dp + \left(\frac{\partial U^s}{\partial V}\right)_{T=0} dV + \left(\frac{\partial U^s_{T=0}}{\partial T}\right)_p dT. \quad (31)$$

The second and fourth terms represent the thermoelastic coupling. The integral of these thermoelastic terms should be equal with opposite sign
$$-\int_{p=0}^{p=p} \left(\frac{\partial U^s_{p=0}}{\partial p}\right)_T dp = \int_{T=0}^{T=T} \left(\frac{\partial U^s_{T=0}}{\partial T}\right)_p dT. \quad (32)$$

The internal energies at zero pressure and at zero temperature are calculated from thermodynamic parameters determined at higher pressure and temperatures. These extrapolations can be considered as a classical approach which does not take into consideration the quantum effects. When sizable quantum effects due to mass, size, and interaction of energy occur then the presented classical description is not applicable.

The internal energy of a system can be determined by integrating Eq. (28) as:

$$U^s = \int_{T=0}^{T=T}\left(\frac{\partial U^s}{\partial T}\right)_p dT + \int_{V_{T,p=0}}^{V_{T,p=p}}\left(\frac{\partial U^s}{\partial V}\right)_T d[V]_T = \int_{T=0}^{T=T}\left(\frac{\partial U^s}{\partial T}\right)_{p=0} dT + \int_{V_{T=0,p=0}}^{V_{T=0,p=p}}\left(\frac{\partial U^s}{\partial V}\right)_{T=0} d[V]_T \quad (33)$$

The four terms given in Eq. (31) are integrated as:

$$\int_{T=0}^{T=T}\left(\frac{\partial U^s}{\partial T}\right)_p dT = \int_{T=0}^{T=T} nc_p dT, \quad (34)$$

and

$$\int_{p=0}^{p=p}\left(\frac{\partial U^s}{\partial p}\right)_T dp = -\int_{V(p=0,T)}^{V(p=p,T)} (p+dp) d[V]_T \cong -\sum_{i=1}^{n}\left(p_i + \frac{\Delta p_i}{2}\right)\Delta[V_i]_T \quad (35)$$

where $p_1=0$ ; $p_i = p_{i-1} + \Delta p_{i-1}$ and $p = p_n + \Delta p_n$,

and

$$\int_{p=0}^{p=p}\left(\frac{\partial U^s_{p=0}}{\partial p}\right)_T dp = \int_{T=0}^{T=T} nc_p dt - \int_{T=0}^{T=T} nc_{p=0} dt, \quad (36)$$

and

$$\int_{p=0}^{p=p}\left(\frac{\partial U^s_{T=0}}{\partial T}\right)_p dT \cong \sum_{i=1}^{n}\left(p_i + \frac{\Delta p_i}{2}\right)\Delta[V_i]_{T=0} - \sum_{i=1}^{n}\left(p_i + \frac{\Delta p_i}{2}\right)\Delta[V_i]_T. \quad (37)$$

The internal energy is a state function which allows testing the proposed thermodynamic description of elastic solids by calculating the internal energy following different paths.

## 3. Testing the equations

MgO has low chemical reactivity and it is stable in large pressure and temperature range which makes it an ideal pressure calibrant. Periclase is the end member of the (Mg,Fe)O solid solution series. Mg-rich ferropericlase is the second most abundant component of the Earth's lower mantle and has significant interest in geophysics. Precise description of the pressure, volume and temperature (p-V-T) relationship of MgO is therefore essential. This importance results in availability of wide range of experimental data and in well defined equation of state (EoS) [8 and ref. there in], which can be ideally used to test the validity of Eqs. (32) and (33). The thermodynamic parameters used in this investigation are given in Table 1.

Using the conventional relationship between the heat capacities the constant pressure heat capacity is calculated as:

$$c_p^s = c_{Debye} + \frac{VTB\alpha^2}{n}. \quad (38)$$

The Debye heat capacity is calculated by using the Debye function [9].

$$c_{Debye} = 3Rf \qquad f = 3\left(\frac{T}{T_D}\right)^3 \int_0^{x_D} \frac{x^4 e^x}{(e^x - 1)^2} dx \quad (39)$$

and

$$x = \frac{h\omega}{2\pi k_B T} \quad \text{and} \quad x_D = \frac{h\omega_D}{2\pi k_B T} = \frac{T}{T_D} \quad (40)$$

where h is the Planck's constant, $\omega$ is the frequency, and $\omega_D$ is the Debye frequency. Equation (39) has to be evaluated numerically [10].

The volume is calculated by using the EoS of Garai [12] which is given as:

$$V = nV_o e^{\frac{-P}{K_o + K_{P1}P + K_{P2}P^2} + (\alpha_o + \alpha_{P1}P + \alpha_{P2}P^2)T + \left(1 + \frac{\alpha_{P1}P + \alpha_{P2}P^2}{\alpha_o}\right)^a \alpha_{T1}T^2} \quad (41)$$

where, $K_{P1}$ is a linear, $K_{P2}$ is a quadratic term for the pressure dependence of the bulk modulus, $\alpha_{P1}$ is a linear and $\alpha_{P2}$ is a quadratic term for the pressure dependence of the volume coefficient of thermal expansion, $\alpha_{T1}$ is a linear term for the temperature dependence of the volume coefficient of thermal expansion and a is constant characteristic of the substance. The theoretical explanations for (41) and the physics of the parameters are discussed in detail [12]. The equation has an analytical solution for the temperature

$$T = \frac{-(\alpha_o + \alpha_{P1}P + \alpha_{P2}P^2) \pm \sqrt{(\alpha_o + \alpha_{P1}P + \alpha_{P2}P^2)^2 + 4\alpha_{T1}\left(1 + \frac{\alpha_{P1}P + \alpha_{P2}P^2}{\alpha_o}\right)^a \left[\ln\left(\frac{V}{V_o}\right) + \frac{P}{K_o + K_{P1}P + K_{P2}P^2}\right]}}{2\alpha_{T1}\left(1 + \frac{\alpha_{P1}P + \alpha_{P2}P^2}{\alpha_o}\right)^a} \quad (42)$$

The pressure can be determined by repeated substitutions as:

$$P = \lim_{n \to \infty} f^n(P) \quad (43)$$

where

$$f^n(P) = \left(K_o + K_{P1}P_{n-1} + K_{P2}P_{n-1}^2\right)\left[\left(\alpha_o + \alpha_{P1}P_{n-1} + \alpha_{P2}P_{n-1}^2\right)T + \left(1 + \frac{\alpha_{P1}P_{n-1} + \alpha_{P2}P_{n-1}^2}{\alpha_o}\right)^a \alpha \right] \quad (44)$$

$n \in \mathbb{N}^*$ and $P_0 = 0$

The convergence of Equation (44) depends on pressure. For the maximum pressure used in this study (up to 140 GPa) n = 15 is sufficient. The maximum convergence error $[\varepsilon]$ for the investigated data set is 0.05 GPa where

$$\varepsilon \geq \left| f^{15}(P) - f^{14}(P) \right|. \quad (45)$$

Using the definition of the bulk modulus [Eq. (13)] the bulk modulus can be determined as:

$$K_{P,T} = \frac{-P}{\ln\left(\frac{V_{P,T}}{V_{P=0,T}}\right)} \quad (46)$$

The volumes in Eq. (46) are calculated by the EoS G, Eq. (41).
Using the EoS G, Eq. (41) the volume coefficient of thermal expansion is calculated as:

$$\alpha = \alpha_o + \alpha_{P1}P + \alpha_{P2}P^2\left(1 + \frac{\alpha_{P1}P + \alpha_{P2}P^2}{\alpha_o}\right)^a \alpha_{T1}T. \tag{47}$$

The Debye temperature in Eq. (39) is the function of pressure and temperature. The Debye temperature for a given temperature and pressure is calculated as:

$$T_D = hk_B^{-1}M^{-\frac{1}{2}}N_A^{\frac{1}{3}}n^{-\frac{1}{3}}V_{mol}^{\frac{1}{6}}B^{\frac{1}{2}} \tag{48}$$

where $k_B$ is the Boltzmann constant, M is the molar mass, $N_A$ is the Avogadro's number, and $V_{mol}$ is the molar volume. In Eq. (48) the bulk sound velocity is calculated as:

$$v_B = \sqrt{\frac{B_T}{\rho}} = \sqrt{\frac{B_T V_{mol}}{M}}. \tag{49}$$

Please note that the bulk modulus calculated by Eq. (46) represents the entire pressure range. The bulk modulus values in Eqs. (48) and (49) are "instantaneous" which can be calculated as:

$$K_{P,T} = \lim_{\Delta P \Rightarrow 0} \frac{-\Delta P}{\ln\left[\frac{V\left(P + \frac{\Delta P}{2}, T\right)}{V\left(P - \frac{\Delta P}{2}, T\right)}\right]}. \tag{50}$$

where the volume is calculated by the EoS G [Eq. (41)]. The difference between the adiabatic and isothermal bulk $[B_T]$ modulus is usually under 1% which will be ignored in this study.

Equation (33) was evaluated at 0, 10, 25, and 50 GPa pressures and 0, 1000, 2000 and 3000 K temperatures. The heat capacities were calculated by using the thermodynamic parameters of MgO and Eqs. (38)-(50). The calculated values fit well to experiments of MgO [13, 14] conducted at ambient pressure [Fig. 1]. The numerical integrations of Eqs. (33) -(37) are given in Table 2. The differences between the internal energies of the two paths are between 0.3-5.4 %, the average is 2.0 %. Based on the complexity of and the approximations used in the calculations the agreement can be considered as good.

## 4. Mechanical equivalency of heat

One of the outcomes of the presented equations is that Joule's law is not universal because only a limited exchange between heat and work is possible in solid phase. Explanation for this limited communication between heat and mechanical energy is proposed here.

In gas phase the energies added to the system increase the kinetic energies of the atoms regardless of their original form of energy. The energy transfers are reversible

$$\text{Heat (thermal energy)} \overset{gas-phase}{\Leftrightarrow} \text{kinetic energy} \tag{51}$$

and

$$\text{Work (mechanical energy)} \overset{gas-phase}{\Leftrightarrow} \text{kinetic energy} \tag{52}$$

resulting in

$$\text{Heat} \overset{gas-phase}{\Leftrightarrow} \text{work}. \qquad (53)$$

Thus the first law of thermodynamics is restored from theoretical considerations.

In solid phase energies from heat and mechanical work are conserved by different physical processes. The thermal energy is conserved by the vibrational energy of the atoms

$$\text{Heat} \overset{solids-phase}{\Leftrightarrow} \text{vibrational energy of the atoms} \qquad (54)$$

The mechanical work is stored by elastic deformation as potential energy

$$\text{Elastic Work} \overset{solids-phase}{\Leftrightarrow} \text{potential energy stored by elastic deformation}. \qquad (55)$$

These energies stored by vibration and elastic deformation are not completely interchangeable. The vibrational frequency of the atoms/molecules changes due to elastic deformation and heat results in volume change which allows a limited energy exchange between heat and work (thermo-elastic coupling) but the majority of heat and work is not transferable in solid phase.

It is suggested that the physics of the mechanical equivalency of heat lies in the energy conservation process. If the thermal and mechanical energies are conserved by the same physical process such as the kinetic energies of the atoms then the mechanical equivalency of heat is valid. However, if the physical process conserving the energies is not the same then thermal and mechanical energies are not completely interchangeable.

If the mechanical work, done on the system, results in plastic deformation or friction then these processes trigger atomic vibrations rather then elastic deformation. In this case the energy of the mechanical work is stored by the same physical process as the thermal energy and the equivalency of work and heat is applicable.

## 5. Conclusions

It is suggested that the contemporary thermodynamic description of elastic solids is incomplete because the isothermal work is not included in the internal energy. It is also suggested that the isobaric work done by the system should not be deducted from the internal energy because the system does not do work on the surrounding when the substance of the surrounding is the same. It is also suggested that Joule's law is not universal and applicable only if the energies from heat and work are stored by the same physical process.

The equations consistent with the proposed thermodynamic description of solids are derived and tested against the experiments of MgO. The calculated internal energies for given pressure and temperature conditions agree well with theoretical predictions, indicating that the proposed thermodynamic description of solids is correct.

The primarily aim of the presented work is to raise awareness that in solid phase there are problems with the fundamental thermodynamic equations. More work has to be done and the proposed thermodynamic description of solids has to be tested on other substances. Consensus, on how the state functions in solid phase should be defined, has to be reached.

**References:**

[1] J.P. Joule, Brit. Assoc. Rep. trans. Chemical Sect., 1845, p. 31.
[2] E. Mach, Die Principien der W¨armelehre, Verlag von J. A. Barth, Leipzig, 1896.
[3] E. Mach, Popul¨ar-Wissenschaftliche Vorlesungen, Verlag von J. A. Barth, Leipzig, 1923.
[4] G. Job, G.; Lankau, T., Ann. N.Y. Acad. Sci. 988 (2003) 171.
[5] J.J. Mares, P. Hubik, J. Sestak, V. Spicka, J. Kristofik, J. Stavek, Thermochimica Acta 474 (2008) 16.
[6] R. Clausius, Annalen der Physik 79 (1850) 368.; 500.; Phil. Mag., 2 (1851) 1.; 102.
[7] L-Y Lu, Y. Cheng, X-R Chen and J. Zhu, Physica B 370 (2005) 236.
[8] J. Garai, J. Chen, G. Telekes, Computer Coupling of Phase Diagrams and Thermochemistry, CALPHAD (2009) accepted.
[9] P. Debye, Ann. Phys. 39 (1912) 789.
[10] Landolt-Bornstein, Zahlenwerte und Funktionen aus Physic, Chemie, Astronomie, Geophysic, und Technik, II. Band, Eigenschaften der Materie in Ihren Aggregatzustanden, 4 Teil, Kalorische Zustandsgrossen, Springer-Verlag, 1961.
[11] J. Garai, A. Laugier, J. Appl. Phys. 101 (2007) 023514.
[12] J. Garai, J. Appl. Phys. 102 (2007) 123506.
[13] I.S. Grigoriev, E.Z. Meilikhov, Handbook of Physical Quantities, CRC Press, Inc. Boca Raton, FL, USA, 1997.
[14] I. Barin, Thermochemical Data of Pure Substances, VCH, Weinheim, Federal republic of Germany, 1989.

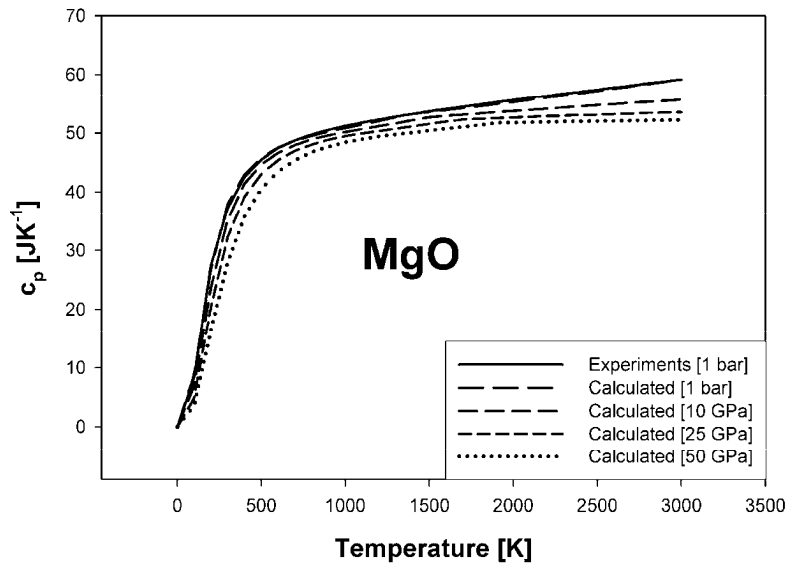

**Fig. 1**. Calculated constant pressure molar heat capacities of MgO at different pressures. Experiments conducted at 1 bar pressure are also plotted

**Table 1.** Thermodynamic parameters of MgO used for the calculations.

| Birch-Murnaghan (p-V-T) EoS [Eq. 41.] | $V_o$ [cm$^3$] | $K_o$ [GPa] | $K_o'$ | $\alpha_o$ [$\times 10^{-5}$ K$^{-1}$] | $\alpha_1$ [$\times 10^{-9}$ K$^{-2}$] | $\delta$ |
|---|---|---|---|---|---|---|
| Periclase [MgO] | 11.137 | 163.59 | 4.145 | 3.197 | 5.784 | 3.690 |

| EoS of Garai [Eq.42] | $V_o$ [cm$^3$] | $K_o$ [GPa] | $\alpha_o$ [$\times 10^{-5}$ K$^{-1}$] | a | b [$\times 10^{-3}$ GPa$^{-1}$] | c [$\times 10^{-7}$ GPa$^{-1}$K$^{-1}$] | d [$\times 10^{-10}$ GPa$^{-2}$K$^{-1}$] | f | g [$\times 10^{-9}$ K$^{-2}$] |
|---|---|---|---|---|---|---|---|---|---|
| Periclase [MgO] | 11.142 | 165.15 | 2.957 | 1.721 | -2.249 | -2.0903 | 4.4 | 10.3 | 6.903 |

**Table 2.** The numerical integrated values of equations 31-35.

| P [GPa] | T [K] | V [Eq. 42] [cm³] | #1 [Eq. 32] [KJ] | #2 [Eq. 33] [KJ] | #3 [Eq. 34] [KJ] | #4 [Eq. 35] [KJ] | #5 [Eq. 31] [KJ] | #6 [Eq. 31] [KJ] | #7 [KJ] | #8 [%] |
|---|---|---|---|---|---|---|---|---|---|---|
| 0 | 0 | 11.138 | | | | | | | | |
| | 1000 | 11.550 | 38.35 | | | | | | | |
| | 2000 | 12.195 | 91.73 | | | | | | | |
| | 3000 | 13.111 | 148.95 | | | | | | | |
| 10 | 0 | 10.549 | | 2.85 | | | | | | |
| | 1000 | 10.866 | 37.00 | 3.29 | 1.36 | -0.45 | 41.20 | 40.29 | 0.91 | 2.2 |
| | 2000 | 11.297 | 89.33 | 4.29 | 2.40 | -1.44 | 94.58 | 93.62 | 0.96 | 1.0 |
| | 3000 | 11.855 | 144.09 | 5.94 | 4.87 | -3.09 | 151.80 | 150.02 | 1.78 | 1.2 |
| 25 | 0 | 9.892 | | 14.14 | | | | | | |
| | 1000 | 10.123 | 35.36 | 16.06 | 2.99 | -1.92 | 52.50 | 51.42 | 1.08 | 2.1 |
| | 2000 | 10.391 | 86.77 | 19.80 | 4.97 | -5.65 | 105.87 | 106.56 | 0.69 | 0.6 |
| | 3000 | 10.697 | 139.91 | 25.63 | 9.05 | -11.48 | 163.09 | 165.53 | 2.44 | 1.5 |
| 50 | 0 | 9.136 | | 42.00 | | | | | | |
| | 1000 | 9.284 | 33.16 | 46.98 | 5.19 | -4.99 | 80.35 | 80.14 | 0.21 | 0.3 |
| | 2000 | 9.436 | 83.52 | 54.89 | 8.22 | -12.89 | 133.73 | 138.40 | 4.67 | 3.4 |
| | 3000 | 9.594 | 135.54 | 66.06 | 13.41 | -24.06 | 190.95 | 201.60 | 10.65 | 5.4 |

#1 $\int_{T=0}^{T=T} c_p^s dT$

#2 $\int_{p=0}^{p=p} \left(\frac{\partial U^s}{\partial p}\right)_T dp = -\int_{V(p=0,T)}^{V(p=p,T)} (p+dp)d[V]_T \cong -\sum_{i=1}^{n} \left(p_i + \frac{\Delta p_i}{2}\right)\Delta[V_i]_T$

where $p_1 = 0$ ; $p_i = p_{i-1} + \Delta p_{i-1}$ and $p = p_n + \Delta p_n$

#3 $\int_{p=0}^{p=p} \left(\frac{\partial U_{p=0}^s}{\partial p}\right)_T dp = \int_{T=0}^{T=T} nc_p dt - \int_{T=0}^{T=T} nc_{p=0} dt$

#4 $\int_{p=0}^{p=p} \left(\frac{\partial U_{T=0}^s}{\partial T}\right)_p dT \cong \sum_{i=1}^{n}\left(p_i + \frac{\Delta p_i}{2}\right)\Delta[V_i]_{T=0} - \sum_{i=1}^{n}\left(p_i + \frac{\Delta p_i}{2}\right)\Delta[V_i]_T$

#5 $U^s = \int_{T=0}^{T=T}\left(\frac{\partial U^s}{\partial T}\right)_{p=0} dT + \int_{p=0}^{p=p}\left(\frac{\partial U^s}{\partial V}\right)_{T=0} d[V]_{T=0}$

#6 $U^s = \int_{T=0}^{T=T}\left(\frac{\partial U^s}{\partial T}\right)_p dT + \int_{p=0}^{p=p}\left(\frac{\partial U^s}{\partial V}\right)_T d[V]_T$

#7 difference between column 5 and 6 in absolute value
#8 percentage of column 7 compared to the average of column 5 and 6.